\documentclass[aps,prl,twocolumn,superscriptaddress,showpacs,preprintnumbers,amsmath,amssymb,floatfix,longbibliography]{revtex4-2}

    \usepackage{graphicx} % Include figure files
    \usepackage{dcolumn} % Align table columns on decimal point
    \usepackage{bm} % bold math
    \usepackage{color} % color
    \usepackage{lipsum} % generates filler text
    \usepackage{graphicx}% Include figure files
    \usepackage{dcolumn}% Align table columns on decimal point
    \usepackage{bm}% bold math
    \usepackage{siunitx} 
    \usepackage{bbold}
    \usepackage{booktabs}
    \usepackage{subfiles}
    \usepackage{python}
    \usepackage{enumitem}
    \usepackage{lineno}
    \usepackage[usenames,dvipsnames]{xcolor}
    \usepackage[linktocpage, colorlinks=true, linkcolor=blue, citecolor=blue, breaklinks=true, urlcolor=blue]{hyperref}
    \usepackage{tikz}
    \usepackage{scalerel}
    \UseRawInputEncoding
    \usetikzlibrary{svg.path}
    \renewcommand{\vec}[1]{\bm{#1}}

    \definecolor{orcidlogocol}{HTML}{A6CE39}
    \tikzset{
      orcidlogo/.pic={
        \fill[orcidlogocol] svg{M256,128c0,70.7-57.3,128-128,128C57.3,256,0,198.7,0,128C0,57.3,57.3,0,128,0C198.7,0,256,57.3,256,128z};
        \fill[white] svg{M86.3,186.2H70.9V79.1h15.4v48.4V186.2z}
                     svg{M108.9,79.1h41.6c39.6,0,57,28.3,57,53.6c0,27.5-21.5,53.6-56.8,53.6h-41.8V79.1z M124.3,172.4h24.5c34.9,0,42.9-26.5,42.9-39.7c0-21.5-13.7-39.7-43.7-39.7h-23.7V172.4z}
                     svg{M88.7,56.8c0,5.5-4.5,10.1-10.1,10.1c-5.6,0-10.1-4.6-10.1-10.1c0-5.6,4.5-10.1,10.1-10.1C84.2,46.7,88.7,51.3,88.7,56.8z};
      }
    }
    \newcommand\orcid[1]{\href{https://orcid.org/#1}{\mbox{\scalerel*{
    \begin{tikzpicture}[yscale=-1,transform shape]
    \pic{orcidlogo};
    \end{tikzpicture}
    }{|}}}}

\begin{document}
    %\linenumbers

% Title and author info
    \title{Collective transport efficiency of microswimmer swarms optimized by tactic run-tumble dynamics}
    
    \author{Maggie Liu\orcid{}}
        \affiliation{Department of Physics \& Astronomy,
        		University of Pennsylvania, Philadelphia, PA 19104}
    \author{Arnold J. T. M. Mathijssen\orcid{0000-0002-9577-8928}}
        \email{amaths@upenn.edu}
        \affiliation{Department of Physics \& Astronomy,
        		University of Pennsylvania, Philadelphia, PA 19104}
    \date{\today}

\begin{abstract}
    The collective motion of microorganisms and microrobots can be used for particle delivery, especially when guided by external magnetic fields, phototaxis, or chemotaxis. This cargo transport is enhanced significantly by hydrodynamic entrainment, where the surrounding fluid and any dissolved molecules or suspended cargo particles are dragged along with a collectively moving swarm. However, it remains unclear how this directed entrainment is affected by stochastic run-tumble motion, and how such motility patterns couple to particle dispersion. Here, we combine theory and simulations to compute the entrainment velocity and diffusivity for different degrees of swimmer directedness. Surprisingly, we find that the transport efficiency P\'{e}clet number, the ratio of advective to diffusive transport, is optimal for intermediate directedness values, so perfectly guided active suspensions perform worse than those with stochastic reorientations. These results could have implications for microrobotic drug delivery and nutrient transport in microbial environments.   
\end{abstract}

%\pacs{Insert relevant PACS numbers here}

\maketitle

\section{Introduction}

Active cargo transport is ubiquitous at the microscale, from the targeted delivery of therapeutics by self-propelled microrobots to nutrient and oxygen redistribution by swimming microorganisms \cite{Nelson2010MicrorobotsMedicine, Elgeti2015PhysicsReview, Bechinger2016ActiveEnvironments, Zottl2016EmergentColloids, Bastos-Arrieta2018BacterialMicroswimmers, Mogre2020GettingWorld, Gompper2025TheRoadmap}. 
In addition to transporting cargo inside their body, these swimmers also drag along their external fluid \cite{Darwin1953NoteHydrodynamics, Pushkin2013FluidMicroswimmersb}, which can contain large amounts of dissolved or suspended cargo particles \cite{Mathijssen2018UniversalCells, Jin2021CollectiveMicroswimmers}.
This entrainment is particularly prominent at low Reynolds numbers, because of the long-rangedness of the hydrodynamic interactions \cite{Pushkin2013FluidMicroswimmersb, Yeomans2014AnMicroorganisms}.
Therefore, hydrodynamic entrainment is fundamental to a vast array of processes, including enhanced diffusion \cite{Wu2000ParticleBath, Kim2004EnhancedBacteria, Leptos2009DynamicsMicroorganisms, Thiffeault2010StirringBodies, Lin2011StirringSquirmers, Morozov2014EnhancedSuspensions, Thiffeault2015DistributionMicroorganisms, Peng2016DiffusionSuspensions, Jeanneret2016EntrainmentObjects, Kanazawa2020LoopySuspensions, Guzman-Lastra2021, Mondal2021StrongMixing, Lin2022StirringSquirming, Kogure2023Flow-inducedSquirmers}, ocean mixing \cite{Leshansky2010DoOcean, Katija2012BiogenicMixing, Houghton2018VerticallyColumn, Ortlieb2019StatisticsMicroswimmers, Aguayo2024FloatingEcosystems, Dabiri2024DoOcean, Barros2025LayeredCarpets}, food uptake \cite{Magar2003NutrientSquirmer, Short2006FlowsTransport, Michelin2011OptimalNumbers, Mathijssen2018NutrientCarpets}, and particle transport \cite{Shum2017EntrainmentInteractions, Mueller2017FluidMicroswimmer, Vaccari2018CargoInterfaces, Purushothaman2021HydrodynamicMicro-channel, Ouyang2023CargoFluid, Lepro2022OptimalMicrocarriers, Liu2025TheSystem}.

Sir Charles G.\,Darwin was the first person to calculate the volume of fluid entrained by a moving particle \cite{Darwin1953NoteHydrodynamics}.
This `Darwin drift volume' diverges for a colloid pulled along an infinite straight line \cite{Eames2003FluidDroplet, Chisholm2017DriftFlows, Shaik2020DragFluid}, but it is finite for a force-free, self-propelled particle \cite{Pushkin2013FluidMicroswimmersb, Chisholm2018PartialSwimmer}.
While a single swimmer may not entrain all that much, a collectively moving school or swarm can displace significant amounts of material \cite{Jin2021CollectiveMicroswimmers}. 

However, these previous works considered either fully directed schools with swimmers moving in exactly the same direction (leading to unidirectional transport), or fully isotropic active baths with swimmers moving in completely random directions (leading to enhanced diffusion).
In reality, the motility patterns of microorganisms guided by environmental stimuli are somewhere in between:
They perform biased random walks \cite{Berg1993RandomBiology} that on average lead to directed motion \cite{Schnitzer1993TheoryChemotaxis}.
\textit{E. coli} bacteria, for example, perform run-tumble dynamics with a chemotactic bias, allowing them to climb nutrient gradients over time \cite{Berg2004E.Motion, Keegstra2022TheChemotaxis}. 
In phototaxis, organisms can steer toward or away from light using different mechanisms: \textit{Chlamydomonas reinhardtii} algae control their angular velocity by coordinating their beating cilia \cite{Polin2009iChlamydomonas/iLocomotion, Bennett2015AiChlamydomonas/i, Arrieta2017PhototaxisReinhardtii, Choudhary2019ReentrantCells}, while \textit{Micromonas} preforms angle-dependent run-tumble dynamics \cite{Henshaw2019PhototaxisCell}.
Other guidance mechanisms include rheotaxis \cite{Mathijssen2019OscillatoryBacteria}, viscotaxis \cite{Liebchen2018iViscotaxis/iGradients}, durotaxis \cite{Sunyer2020Durotaxis}, galvanotaxis \cite{Nwogbaga2023PhysicalGalvanotaxis}, and magnetotaxis \cite{Waisbord2021FluidicEnvironments}, to name a few, which can also be used to control microrobots \cite{Zhou2021MagneticallyNanorobots}.  
It remains unknown how these directed yet stochastic dynamics affect the collective transport and dispersion of cargo particles in microswimmer swarms.

In this work, we address this question by developing a comprehensive model for collective cargo transport by tactic active suspensions, combining biased run-and-tumble kinematics with entrainment hydrodynamics. Through a combination of analytical theory and large-scale numerical simulations, we systematically investigate how a tunable tactic strength, $\alpha$, governs both the mean drift and the anisotropic diffusion of passive tracers in these guided microswimmer swarms. We reveal a non-trivial and non-monotonic relationship between these transport properties. Our central finding is that the efficiency of directed transport is not maximized by the strongest possible directional bias. Instead, there exists an optimal, intermediate bias that balances the need for rapid, directed travel with the requirement of minimal dispersion before reaching the delivery target. 

%%%%%%%%%%%%%%%%%%%%%%%%%%%%%%%%%%%%
%%%%%%%%%%%%%%%%%%%%%%%%%%%%%%%%%%%%
\section{Model for directed swimmer motion}
\label{sec:SwimmerModel}

\begin{figure*}[t]
    \centering
    \includegraphics[width=1\linewidth]{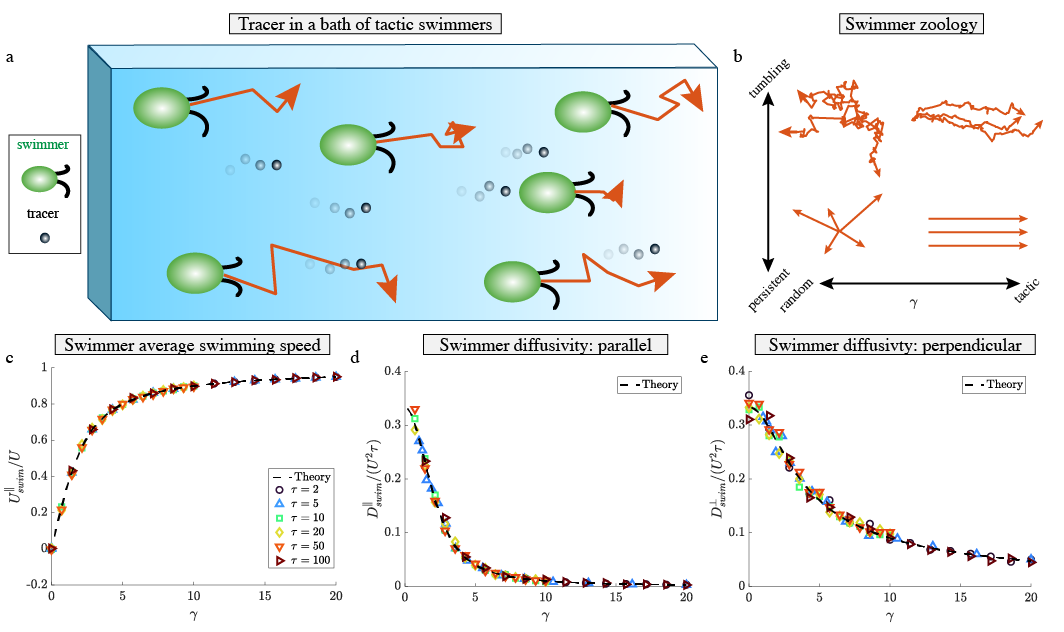}
    \caption{
        \label{fig:1}
        \textbf{Directed microswimmer swarms.}
        %CAN WE ALSO MAKE THIS 2 ROWS INSTEAD OF 3, JUST LIKE FIGURE 2?
        \textbf{(a)} Overview of the model. Tracers are advected by a guided active suspension, leading to both mixing and directed transport. 
        \textbf{(b)} Representative swimmer trajectories as a function of the tactic strength $\gamma$ and the inverse run time $1/\tau$, illustrating a range of behaviors from random to highly tactic motion. 
        \textbf{(c)} Mean swimmer velocity as a function of $\gamma$ for different run times $\tau$, as predicted by Eq.~(\ref{eq:SwimmerDirectedness}). 
        \textbf{(d)} Swimmer diffusivity in the direction parallel to the swarm, as a function of $\gamma$ and $\tau$. The dashed line corresponds to the theoretical prediction of Eq.~(\ref{eq:SwimmerDiffusivityZ}). 
        \textbf{(e)} Swimmer diffusivity perpendicular to the swarm. The dashed line corresponds to the theoretical prediction of Eq.~(\ref{eq:SwimmerDiffusivityXY}). 
    }
\end{figure*}

%%%%%%%%%%%%%%%%%%%%%%%%%%%%%%%%%%%%
\subsection{Biased run-tumble dynamics}
\label{subsec:SwimmerModel}

We consider a dilute suspension of identical microswimmers in a three-dimensional fluid, as depicted schematically in Figure \ref{fig:1}a.
The preferred direction of motion is along an external field or gradient, such as a magnetic field for magnetotaxis, a nutrient gradient for chemotaxis, or light for phototaxis.
Without loss of generality, we define this preferred direction to be along $\hat{\vec{z}}$, using standard Cartesian coordinates.
The swimmers are spherical with radius $R$, and they move with a constant speed $U$ in their orientation direction $\hat{\vec{p}}$. 
The number density of the swimmers is $n$, and we consider the dilute limit $(nR^3\ll1)$ without swimmer-swimmer interactions.

The motion of each swimmer is described with directed run-and-tumble dynamics, a biased random walk that captures the essential motility features of organisms such as \textit{Chlamydomonas} algae or \textit{E. coli} bacteria: 
During the `run' phase, the swimmer travels in a straight line segment along $\hat{\vec{p}}$. 
The duration of each run is a random variable drawn from an exponential distribution, $f_{\tau}(t_\textmd{run})$, giving rise to a Poisson process with a characteristic run time $\tau$.
These runs are punctuated by instantaneous `tumble' events, where the swimmer adopts a new orientation that is chosen from the three-dimensional von Mises-Fisher distribution, 
\begin{equation}
    \label{eq:VonMisesFisher}
    f_{\gamma}(\hat{\vec{p}})
    = \frac{\gamma}{4\pi\sinh(\gamma)}\exp(\gamma \, \hat{\vec{p}} \cdot \hat{\vec{z}}),
\end{equation}
which looks like a 2D Gaussian wrapped around a spherical shell (2-sphere). 
Here $\gamma$ is the tactic strength, a parameter that controls how much the organisms respond to the external signal. 

Figure~\ref{fig:1}b illustrates this with representative swimmer trajectories from our simulations.
When $\gamma=0$, all orientations are equally likely [left side], whereas in the limit $\gamma \rightarrow \infty$, the swimmers move along $\hat{\vec{z}}$ [right].
When the run time $\tau$ is small, the swimmers often tumble [top], and when it is large, the swimmers move in straight lines [bottom].

%%%%%%%%%%%%%%%%%%%%%%%%%%%%%%%%%%%%
\subsection{Swimmer directedness}
\label{subsec:SwimmerDirectedness}

Next, we quantify how quickly the swarm moves on average along the preferred direction. 
The mean velocity of the swimmers is given by the first moment \cite{Hillen2017MomentsApplications} of the orientation distribution [Eq.~\ref{eq:VonMisesFisher}], that is
	\begin{equation}
		\langle\vec v^\textmd{swimmers} \rangle = U \langle \hat{\vec{p}} \rangle = U \int_{\mathbb{S}^2} f_{\gamma}(\hat{\vec{p}})\, \hat{\vec{p}} \, d\hat{\vec{p}}.
	\end{equation}
By symmetry, this simplifies using 
$\langle \hat{\vec{p}} \rangle = \langle p_z \rangle \hat{\vec{z}} = \alpha \hat{\vec{z}}$.
	% \begin{equation}
	% 	\langle \hat{\vec{p}} \rangle = \langle p_z \rangle \hat{\vec{z}} = \alpha \hat{\vec{z}},
	% \end{equation}
Here we defined the swimmer directedness $\alpha = \langle p_z \rangle$, which can be integrated analytically to yield
	\begin{align}
        \label{eq:SwimmerDirectedness}
		\alpha &= \coth(\gamma)-1/\gamma.
	\end{align}
    
This result agrees well with direct simulations of swimmers performing run-tumble motion [Fig.~\ref{fig:1}c].
The swimmer directedness $\alpha$ increases monotonically from zero to one as the tactic strength $\gamma$ increases from zero to $\infty$. Notably, the directedness is independent of the run time, validated for a broad range of $\tau$ values [colours].

%%%%%%%%%%%%%%%%%%%%%%%%%%%%%%%%%%%%
\subsection{Swimmer dispersion}
\label{subsec:SwimmerDispersion}

After determining the average velocity of the swarm, we consider how quickly it spreads.
The diffusivity tensor of the swimmers is described by the Green-Kubo formula \cite[see e.g.][]{Shalchi2011ApplicabilityTheory}, given by
	\begin{align}
        \label{eq:Green-Kubo}
		D_{ij}=\int_0^{\infty}\langle\delta v_i(t)\delta v_j(0)\rangle\,dt,
	\end{align}
where $\delta\vec v(t)=\vec v(t)-\vec v(0)$ is the swimmer velocity fluctuation at time $t$ compared to $t=0$, noting that we dropped the superscript `swimmers' for brevity. 
The velocity autocorrelation function (VACF) is given by 
	\begin{align}
        \label{eq:VACF}
		\langle\delta\vec v(t)\cdot\delta\vec v(0)\rangle
        = \langle|\delta\vec v(0)|^2\rangle e^{-t/\tau},
	\end{align}
where
	\begin{align}
   		\langle |\delta\vec v(0)|^2\rangle
        = &\langle|\vec v(0)-\langle \vec v\rangle|^2\rangle\\
        = &\langle|U\hat{\vec{p}} - \alpha U \hat{\vec{z}}|^2\rangle \\
        % =&U^2\langle(\hat{\vec{p}}-\alpha\hat{\vec{z}})\cdot(\hat{\vec{p}}-\alpha\hat{\vec{z}})\rangle\\
        =&U^2\langle|\hat{\vec{p}}|^2-2\alpha(\hat{\vec{p}}\cdot\hat{\vec{z}})+\alpha^2|\hat{\vec{z}}|^2\rangle\\
        =&U^2(1-\alpha^2).
	\end{align} 
Due to the axial symmetry of the reorientation distribution around the $\hat{\vec{z}}$ axis, the diffusivity tensor must be diagonal, with components $D_{xx}=D_{yy}=D_\perp$ and $D_{zz}=D_\parallel$. 
For convenience, we first compute the variance of $\hat{\vec{p}}$: 
Since $\langle \hat{p}_z^2 \rangle = 1 - \frac{2}{\gamma}\left(\coth\gamma - \frac{1}{\gamma}\right) = 1 - \frac{2\alpha}{\gamma}$, the variance of the parallel component is
    \begin{align}
    \label{eq:VarianceParallel}
    \mathrm{Var}(\hat{p}_z) = \langle \hat{p}_z^2 \rangle - \langle \hat{p}_z \rangle^2 = \left(1 - \frac{2\alpha}{\gamma}\right) - \alpha^2.
    \end{align}
Correspondingly, the perpendicular component is given by $\langle p^2_x\rangle=\frac{1}{2}(1-\langle p_z^2\rangle)=\frac{\alpha}{\gamma}$, and $\langle p_x\rangle=0$. 
Now we can integrate the VACF components to find the diffusivity
    \begin{align}
    \label{eq:SwimmerDiffusivityZ}
    D_\parallel=&\int_0^\infty\langle\delta v_z(t)\delta v_z(0)\rangle\, dt\\=&U^2\mathrm{Var}(\hat p_z)\int_0^\infty e^{-t/\tau}\, dt\\=& U^2\tau\left(1-\frac{2\alpha}{\gamma}-\alpha^2\right),
    \end{align}
and 
    \begin{align}
    \label{eq:SwimmerDiffusivityXY}
    D_\perp=U^2\tau\frac{\alpha}{\gamma}.
    \end{align}
    
Figures \ref{fig:1}d-e compare this theory for the swimmer diffusivity tensor with simulations.  
In the limit of no tactic bias ($\gamma \rightarrow 0$), we can use a Taylor expansion to approximate $\alpha\approx\gamma/3$. 
The parallel diffusivity then becomes $D_\parallel\rightarrow U^2\tau/3$, and the perpendicular diffusion also becomes $D_{\perp}\rightarrow U^2\tau/3$. 
In the limit of a strong bias ($\gamma\rightarrow\infty$), the parallel diffusivity vanishes, $D_\parallel\rightarrow 0$, as the motion becomes almost deterministic along the $\hat{\vec{z}}$ axis and the fluctuations disappear. 
The perpendicular diffusivity also vanishes, $D_\perp\rightarrow 0$, as the motion perpendicular to the gradient is suppressed.
Both diffusivities $D_\parallel$ and $D_\perp$ do depend on the run time $\tau$, but after rescaling by $\tau U^2$, they nicely collapse onto the same curves.

%%%%%%%%%%%%%%%%%%%%%%%%%%%%%%%%%%%%
%%%%%%%%%%%%%%%%%%%%%%%%%%%%%%%%%%%%
\section{Collective cargo transport}

\begin{figure*}[t]
   \centering
   \includegraphics[width=1\linewidth]{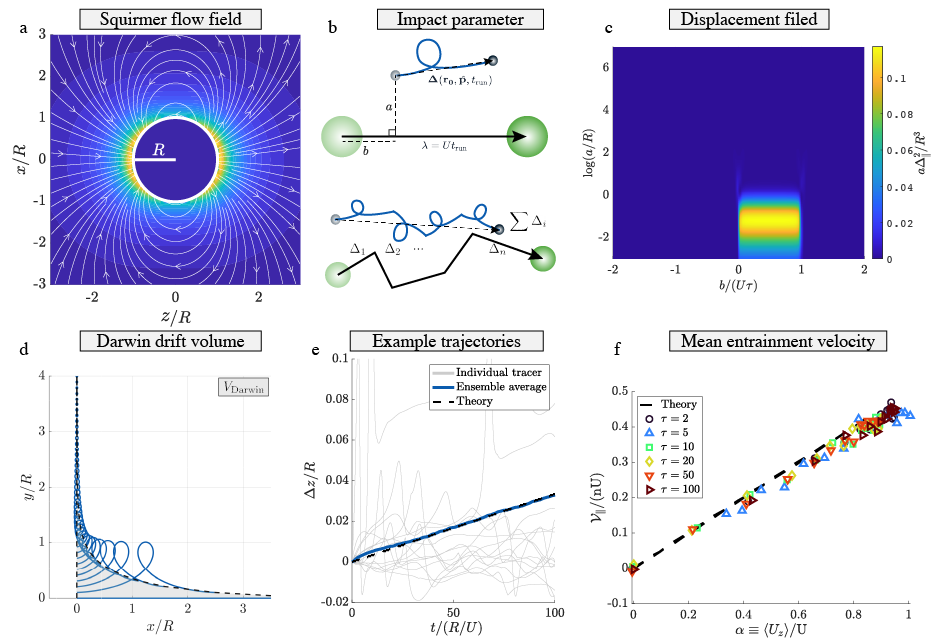}
    \caption{
    \label{fig:2}
    \textbf{Transport by phototactic swimmers.} 
     \textbf{(a)} The flow field of a swimmer with radius $R$ is characterized by the neutral squirmer model. The flow field resembles a source dipole in the far field. 
    \textbf{(b)} The impact parameter $a,b$ is the perpendicular and tangential component of the separation vector at the beginning of a run, immediately following a tumbling event. 
    \textbf{(c)} Displacement vector $\Delta$ as a function of impact parameters. The majority of displacements occur when a swimmer eclipses the tracer during its swimming path. %I changed the order of c and d
    \textbf{(d)} Darwin drift volume is defined by the net volume enclosed by the displacement of an imaginary film when swimmer start's from infinity far behind and travel to infinity far beyond. 
    \textbf{(e)} Net tracer displacement grows linearly with time. The gray line represents randomly selected individual trajectories. Simulation results collapse onto the line $n U V_\textmd{Darwin}$ [Eq.~\ref{eq:EntrainmentVelocity}]. 
    \textbf{(f)} Normalized entrainment speed shown as a function of phototaxis strength $\alpha$ agrees with prediction.
    }

\end{figure*}

After considering the velocity and dispersion statistics of the swimmers, we focus on tracer particles that are dragged along by the swimmers [Fig.~\ref{fig:1}a; grey] because of the long-ranged flows they generate at low Reynolds number \cite{Pushkin2013FluidMicroswimmers, Jin2021CollectiveMicroswimmers}.
Note that these tracers can either represent a solid particle or a fluid parcel. 
In other words, each swimmer can entrain a substantial fluid volume, which is called the Darwin drift volume, $V_\textmd{Darwin}$ \cite{Darwin1953NoteHydrodynamics}.
When a swarm of swimmers moves in one direction, the tracers will be displaced multiple times by subsequent entrainment events. 
Thus, the fluid (and everything in it) will be dragged along with a drift velocity $\mathcal{V}$ that depends on the swimmer velocities $\vec{v}$, their number density $n$, and their individual drift volume $V_\textmd{Darwin}$. 

%%%%%%%%%%%%%%%%%%%%%%%%%%%%%%%%%%%%
\subsection{Swimmer-generated flows}
\label{subsec:SwimmerFlows}

To determine this entrainment velocity, we must model the flow field produced by each swimmer. 
The hydrodynamics of swimming microorganisms are complex, but at distances larger than the organism's size, it can often be approximated by the leading-order terms in a multipole expansion of the Stokes equation \cite[see e.g.][]{Mathijssen2015TracerSurface}. 
The squirmer model is a canonical framework for this, representing a spherical swimmer propelled with a surface slip velocity \cite{Blake1971APropulsion, Lauga2020TheMotility}.
Here, we consider a neutral squirmer whose flow is given by 
%\textcolor{red}{far-field velocity is that of a source dipole,}
%\textcolor{blue}{flow field is given by}
    \begin{align}
    \label{eq:SquirmerModel}
    \vec u(\vec r)
  = \frac{B_{1}R^{3}}{r^{3}}
     \left[ \left(\hat{\vec r} \cdot \hat{\vec{p}} \right) \hat{\vec r}
            - \frac{1}{3} \hat{\vec{p}} \right] \qquad r \ge R,
    \end{align}
where $\vec{r} = \vec{x} - \vec{x}_\textmd{sw}$ is the relative distance vector between the swimmer position and the point where the flow is evaluated. 
This flow field of a neutral squirmer is depicted in figure~\ref{fig:2}a.
The far-field velocity corresponds to that of a source dipole: 
Fluid is drawn in along the swimmer's equator and expelled symmetrically along its swimming axis. This structure arises from the leading-order term in the squirmer expansion (the $B_1$ mode). The streamlines exhibit a fore-aft symmetry, producing no long-ranged Stokeslet component.

%%%%%%%%%%%%%%%%%%%%%%%%%%%%%%%%%%%%
\subsection{Tracer displacement for a single swimmer run}
\label{subsec:TracerDisplacement}

Next, we consider how these swimmer-generated flows can displace tracer particles [Fig.~\ref{fig:2}b].
Interestingly, when a single force-free swimmer moves along an infinite straight line, a stationary point in the fluid would experience zero net displacement due to the front-back symmetry of the flow \cite{Mathijssen2015TracerSurface}. 
However, a tracer is not stationary; it is a Lagrangian particle that is advected by the very flow it is in: 
Because the tracer moves, it does not sample the forward- and backward-pushing regions of the flow field for equal amounts of time. 
Consequently, the tracer is first pushed forwards, then it moves around a loop, and then it is drawn forwards again. 
This broken symmetry leads to a net tracer displacement. 
% Especially the particles that come close to the swimmer can be dragged along for large distances [Fig.~\ref{fig:2}b]. 

We quantify the tracer displacement following the framework from \citet{Lin2011StirringSquirmers}. 
The total tracer displacement vector for a single swimmer run is given by 
    \begin{align}
    \label{eq:Displacement}
    \vec{\Delta}(\vec{r_0},\hat{\vec{p}},t_\textmd{run}) = \int_0^{t_\textmd{run}}\vec{u}\left(\vec{x}_\textmd{tr}(t)-\vec{x}_\textmd{sw}(t); ~ \hat{\vec{p}}\right)\,dt,
    \end{align}
where $t_\textmd{run}$ is the duration of the swimmer run, and $\vec{r}_{0}=\vec{x}_{tr}(0)-\vec{x}_{sw}(0)$ is the initial tracer-swimmer separation vector. This $\vec{\Delta}$ encapsulates the hydrodynamic ``kick'' imparted onto the tracer. 

The displacement imparted by the swimmer on the tracer depends on the impact parameters $a$ and $b$, as shown figure \ref{fig:2}b. In general, $\vec{\Delta}$ is a function of initial swimmer-tracer separation, the swimming direction, and the run duration. Over many tumbling events, the net displacement can be approximated as the sum of individual contributions.

For infinite swimming lines ($t_\textmd{run}\rightarrow \infty$), the Darwin drift volume is equal to the grey-shaded region in figure \ref{fig:2}d, given by
    \begin{align}
    \label{eq:DarwinDrift}
    V_\textmd{Darwin} = \int_0^\infty 2 \pi a \Delta_z(a, \hat{\vec{z}},\infty) da,
    \end{align}
%CHECK/UPDATE RHO_0 HERE.
where $a$ is the initial distance between the tracer and the swimming line, using cylindrical coordinates.
For a neutral squirmer of radius $R$, this drift volume is 
$V_\textmd{Darwin} = \frac{2}{3}\pi R^3$,
    % \begin{align}
    % \label{eq:DarwinDriftSquirmer}
    % V_\textmd{Darwin} = \frac{2}{3}\pi R^3,
    % \end{align}
which is half of the swimmer volume \cite{Pushkin2013FluidMicroswimmers}.
Note that besides the squirmer model, more detailed models have been developed that account for effects like flagellar motion \cite[see e.g.][]{Mathijssen2018UniversalCells}, but for these models the Darwin drift volume remains unknown.

%%%%%%%%%%%%%%%%%%%%%%%%%%%%%%%%%%%%
\subsection{Tracer transport by multiple entrainment events}
\label{subsec:TracerTransport}

We can now build upon this single-run concept to understand the mean entrainment velocity for tracers in a swarm of phototactic swimmers. 
Figure \ref{fig:2}e shows trajectories of tracer particles that are pushed forwards multiple times by successive swimmer encounters.
During the encounters, they experience large displacements events, and between encounters, they experience comparatively weak Brownian motion and long-ranged flows.
Averaged over time, or equivalently over a statistical ensemble of tracers, these dynamics give rise to a mean entrainment velocity, 
	$\mathcal{V} = \langle \partial \vec{x}_\textmd{tr} / \partial t \rangle.$
This ensemble-averaged transport is shown by the blue line. 

To understand this from a theory point of view [dashed line], we note that a single swimmer with mean velocity $U \hat{\vec{p}}$ entrains a volume of fluid $V_\textmd{Darwin}$, resulting in volumetric flux of $(U\hat{\vec{p}})V_\textmd{Darwin}$. 
The total entrainment stems from the sum of contributions from all swimmers, giving us the mean entrainment velocity
\begin{align}
    \mathcal{V}_\parallel
    &= n UV_\textmd{Darwin}\langle \hat{\vec{p}} \rangle
    \nonumber \\
    \label{eq:EntrainmentVelocity}
    &\approx \frac{2}{3}\pi a^3 n U (\coth(\gamma)-1/\gamma) \hat{\vec{z}},
\end{align}
Where we combined Eq.~\ref{eq:SwimmerDirectedness} for the swimmer directedness $\langle \hat{\vec{p}} \rangle$ and Eq.~\ref{eq:DarwinDrift} for the drift volume $V_\textmd{Darwin}$.

Figure~\ref{fig:2}f plots the normalized entrainment velocity as a function of the tactic strength $\alpha$. 
The data points, representing simulations with a wide range of run times $\tau$, collapse onto a single straight line passing through the origin.
While the prediction of Eq.~\ref{eq:EntrainmentVelocity} uses the drift volume for an infinitely long swimming line, the simulations agree remarkably well with this approximation, even for frequently tumbling swimmers with short swimming paths.
Thus, the entrainment velocity is independent of the individual swimmer persistence length and depends only on the collective directional bias $\alpha$.

\section{Tracer Dispersion}

\begin{figure*}[t]
   \centering
   \includegraphics[width=0.8\linewidth]{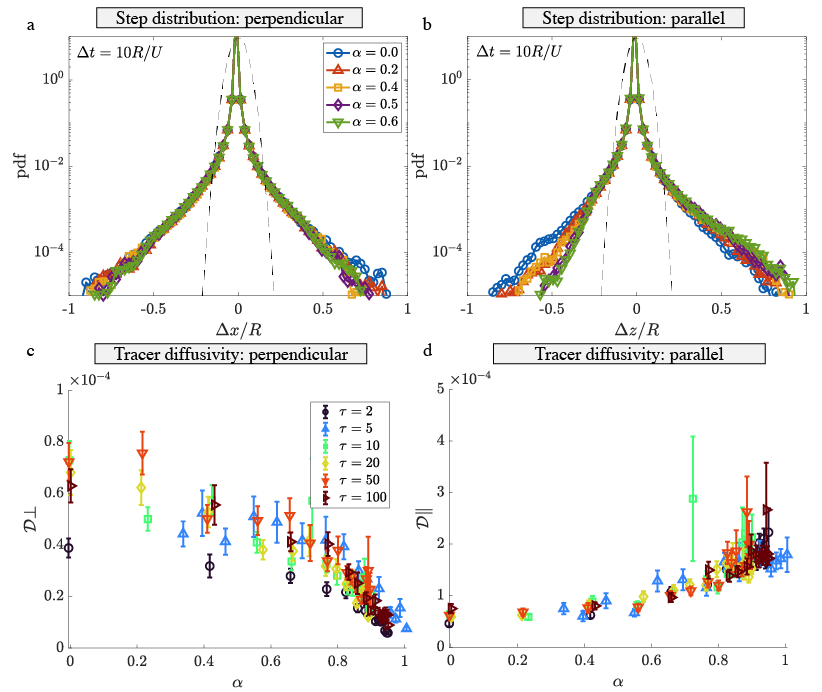}
    \caption{
    \label{fig:3}
    \textbf{Mixing induced by collectively moving swimmers.} 
    \textbf{(a,b)} Probability distribution functions (PDFs) of tracer displacements in the perpendicular and parallel directions, during a time interval $\Delta t = 10R/U$.
    The former is symmetric, but the latter is not, leading to a non-zero mean displacement.
    \textbf{(c,d)} The effective tracer diffusivity in the perpendicular and parallel direction.  $\mathcal{D}_{\perp}$ decreases as a function of $\alpha$, while $\mathcal{D}_{\parallel}$ increases.
    }
\end{figure*}

After computing the mean tracer velocity, we consider their diffusivity due to multiple entrainment events.
Unlike previous studies on enhanced diffusion due to swimmer mixing in isotropic active baths, we must consider anisotropic diffusion due to the directed swarm motion: 
analogous to the swimmer diffusivity [Eq.~\ref{eq:SwimmerDiffusivityZ}-\ref{eq:SwimmerDiffusivityXY}], the tracer diffusivity tensor $\mathcal{D}_{ij}$ is diagonal with entries $\mathcal{D}_\parallel$ and $\mathcal{D}_\perp$ in the parallel and perpendicular directions.
Note the distinction in notation between $D$ for the swimmer diffusivity and $\mathcal{D}$ for the tracer diffusivity. 

In the dilute limit, swimmer-tracer encounters are sparse, and can thus be treated as independent scattering events \cite{Kanazawa2020LoopySuspensions, Thiffeault2015DistributionMicroorganisms}. 
We can compute the long-time drift of tracer as the average displacement per encounter times the encounter rate. We sum over all possible swimmer runs, which yields the tracer diffusivity tensor 
\begin{align}
    \label{eq:TracerDiffsivity}
    \mathcal{D}_{ij}=\frac{n}{2}\langle\dot N_\textmd{enc}\rangle\langle(\Delta_i-\langle\Delta_i\rangle)(\Delta_j-\langle\Delta_j\rangle)\rangle_\textmd{enc},
\end{align}
where $\langle \dot N_\textmd{enc}\rangle$ is the average rate of encounters, and the second term is the covariance of displacement vector averaged over all encounter parameters. 
For our run-and-tumble swimmers, the rate of new encounters is equal to the tumble rate, $1/\tau$. This leads to the expression for the parallel component of diffusivity: 
\begin{align}
    \label{eq:TracerDiffsivityZ}
    \mathcal{D}_\parallel
    =\frac{n}{2\tau}\left\langle(\Delta_z(\vec r_0,\hat{\vec{p}},t_\textmd{run})-\langle\Delta_z\rangle)^2\right\rangle_{\vec r_0,\hat{\vec{p}},t_\textmd{run}}.
\end{align}
This integral is analytically intractable, so we proceed numerically. 

Figure \ref{fig:3}a shows the probability distribution function (PDF) of perpendicular tracer displacements during a time interval $\Delta t=10R/U$.
Consistent with previous studies, these PDFs are not Gaussian but feature long tails \cite{Leptos2009DynamicsMicroorganisms}. 
Moreover, they are symmetric in the $x$ and $y$ directions.
In contrast, figure \ref{fig:3}b shows the PDFs for the parallel tracer displacements, $\Delta_z$. These functions are increasingly more asymmetric as $\alpha$ increases, leading to a non-zero entrainment velocity. 
Figures \ref{fig:3}c,d show the corresponding tracer diffusivities $\mathcal{D}_\perp$ and $\mathcal{D}_\parallel$ from our simulations.
Surprisingly, these quantities behave very differently as $\alpha$ increases:

The perpendicular component decreases monotonically with $\alpha$. This is because a stronger bias aligns the swimmers more closely with the $z$-axis. Consequently, their motion in the transverse plane is suppressed, and the hydrodynamic kicks they impart to the tracer are predominantly in the parallel direction. This is consistent with the swimmer diffusivity $D_\perp$ in figure \ref{fig:1}e. 

However, the parallel tracer diffusivity increases strongly as $\alpha$ increases. 
As swimmers are more aligned, the hydrodynamic displacements they cause in the $z$ direction become more correlated and persistent.  
The increased coherence of the swimmer paths leads to larger variance in the forward `kicks,' resulting in a significantly enhanced longitudinal diffusion. This is much unlike the trend of $D_\parallel$ in figure \ref{fig:1}d and rather surprising.

%By combining the entrainment velocity [Fig.~\ref{fig:2}] with the diffusivity [Fig.~\ref{fig:3}], we arrive at the complete coarse-grained transport equation for the concentration of tracer particles, $C(\vec x,t)$. 
%This advection-diffusion equation is given by
%\begin{align}
%    \label{eq:AdvectionDiffusionEqn}
%    \frac{\partial C}{\partial t}+\nabla\cdot( \mathcal{V} C)=\nabla\cdot(\mathcal{D} \nabla C).
%\end{align}

\section{Optimal Transport Efficiency}

\begin{figure*}[t]
\centering 
\includegraphics[width=0.6\linewidth]{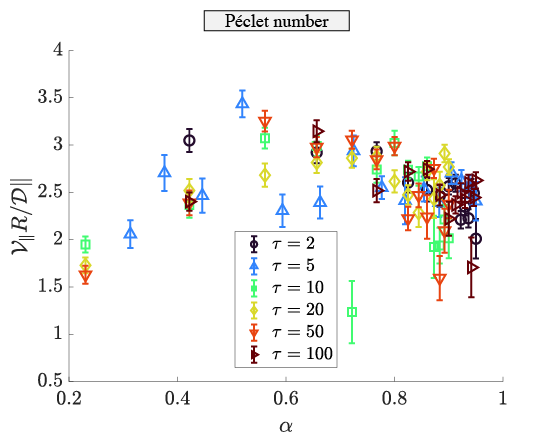}
\caption{
    \label{fig:4}
    \textbf{Transport efficiency P\'{e}clet number.}
    The ratio of transport to mixing reaches a maximum at an intermediate tactic strength, so the most efficient transport occurs around $\alpha=0.6$.
    }
\end{figure*}

Until now, we have characterized the drift and diffusion separately, but a practical measure of transport effectiveness must consider both. 
For applications like nutrient foraging or drug delivery, the goal is not just to move quickly but to deliver a payload to a target region without excessive dispersion. 
That is, a high entrainment velocity is of little use if the associated diffusion is so large that the concentration at the target becomes negligible.

To quantify this balance, we introduce a dimensionless tracer P\'{e}clet number which compares the rate of advective transport to the rate of diffusive spreading in the swarm direction of motion:
\begin{align}
    \label{eq:Peclet}
    \text{Pe}_\textmd{tracers}=\frac{a \mathcal{V}_\parallel}{\mathcal{D}_{\parallel}}.
\end{align}
Figure \ref{fig:4} shows this transport efficiency ratio as a function of the swimmer directedness $\alpha$. 

As before, the data for all persistence times, $\tau$, spanning two orders of magnitude, collapse onto a single universal curve. This implies that the efficiency of hydrodynamically driven transport is governed solely by the collective directional bias of population, $\alpha$, and is independent of the microscopic details of individual swimmer paths. 

Finally, our central result: 
The transport efficiency is a non-monotonic function of the swimmer directedness. 
It is zero for isotropic swimmers ($\alpha=0$), as there is no net transport. 
The efficiency then increases with $\alpha$, reaching a maximum at an intermediate bias of approximately $\alpha\approx0.6$. 
Beyond this peak, as the swimmer motion becomes nearly ballistic ($\alpha\rightarrow1$), the efficiency surprisingly decreases. 

This result is explained by the strong non-monotonic increase of the tracer diffusivity [Fig.~\ref{fig:3}d] compared to the linear increase of the entrainment velocity [Fig.~\ref{fig:3}f]. At low $\alpha$, $\mathcal{D}_{\parallel}$ is almost constant so the P\'{e}clet number increases with $\mathcal{V}_\parallel$. At high $\alpha$, the diffusivity increases rapidly and dominates the transport efficiency.

\section{Discussion}
%SUMMARIZE OUR KEY RESULTS. ONE SENTENCE FOR EACH FIGURE, APPROXIMATELY, ENDING WITH THE MAIN RESULT.
In this work, we develop a comprehensive model for tactic swimmers and calculate their corresponding swim speed and diffusivity. Using this model, we predicted and verified the entrainment velocity of passive tracer in a suspension of tactic swimmers. We further characterized the diffusive properties and dispersion pattern of the tracer as a function of tactic strength. Our results demonstrate the existence of an optimal transport efficiency, which emerges at intermediate tactic strengths.
The most effective transport is achieved not by the most directed swimmers, but by a population that retains a specific degree of randomness in its motion. This is a form of stochastic resonance \cite{McDonnell2009WhatBiology, Sagues2007SpatiotemporalNoise}, where small amounts of orientational noise can be beneficial. 

Given the fundamental tradeoff between exploration and directed motion, our result shows the peak in the transport efficiency may represent an evolutionarily optimized solution \cite{Biswas2023ModeProblems}. %ADD CITATIONS FOR THIS.
The swimmer maintains a strong directional bias, allowing them to reach the promising region quickly and efficiently. 
However, they retain just enough randomness to ensure that once they arrive, they can effectively mix and sample the local environment, maximizing their chances of encountering dispersed resources.  

Future work should extend this model to explore the effects of higher swimmer concentrations, where direct hydrodynamic interactions become important and can lead to complex collective phenomena \cite{Mathijssen2019CollectiveWaves, Furukawa2014Activity-inducedInteractions}. %ADD CITATIONS FOR THIS.
Investigating different swimmer types, such as pushers and pullers, will also be crucial, as their interactions are known to produce qualitatively different collective states \cite{Bardfalvy2020SymmetricSuspensions}. %ADD CITATIONS FOR THIS. 
Finally, experimental verification of these theoretical predictions using phototactic microorganisms or synthetic active colloids would be a valuable next step in understanding the design rules for natural and microrobotic cargo delivery systems.

%%%%%%%%%%%%%%%%%%%%%%%%%%%%%%%%%%%%%%%%%%%%%%%%%%
\section*{Acknowledgements}
%%%%%%%%%%%%%%%%%%%%%%%%%%%%%%%%%%%%%%%%%%%%%%%%%%
We are grateful to all members of the Mathijssen lab for helpful discussions and critical feedback. We acknowledge funding from the National Science Foundation (UPenn MRSEC DMR-2309043) and the Charles E. Kaufman Foundation (Early Investigator Research Award KA2022-129523; New Initiative Research Award KA2024-144001).

%%%%%%%%%%%%%%%%%%%%%%%%%%%%%%%%%%%%%%%%%%%%%%%%%%
\section*{Author contributions}
%%%%%%%%%%%%%%%%%%%%%%%%%%%%%%%%%%%%%%%%%%%%%%%%%%
A.M. designed the research and provided funding. M.L. developed the analytical theory, performed the simulations, and made the figures. Both authors analyzed the data and wrote the manuscript.

%%%%%%%%%%%%%%%%%%%%%%%%%%%%%%%%%%%%%%%%%%%%%%%%%%
\section*{Competing interests}
%%%%%%%%%%%%%%%%%%%%%%%%%%%%%%%%%%%%%%%%%%%%%%%%%%
The authors declare no competing interests.

%%%%%%%%%%%%%%%%%%%%%%%%%%%%%%%%%%%%%%%%%%%%%%%%%%
\section*{Data availability}
%%%%%%%%%%%%%%%%%%%%%%%%%%%%%%%%%%%%%%%%%%%%%%%%%%
The data supporting the findings of this article are openly available at \cite{DataAvailability2025}.

%%%%%%%%%%%%%%%%%%%%%%%%%%%%%%%%%%%%%%%%%%%%%%%%%%
\section*{Code availability}
%%%%%%%%%%%%%%%%%%%%%%%%%%%%%%%%%%%%%%%%%%%%%%%%%%
The simulation code used in this paper is available from \url{https://github.com/yanagi814/entrainment-stirring}

%%%%%%%%%%%%%%%%%%%%%%%%%%%%%%%%%%%%%%%%%%%%%%%%%%
\section*{Appendix: Simulation Details}
%%%%%%%%%%%%%%%%%%%%%%%%%%%%%%%%%%%%%%%%%%%%%%%%%%
All simulations and subsequent data analysis were performed using custom scripts written in MATLAB R2024b. The simulated system consists of $N_\textmd{swim}$ swimmers and a single passive tracer particle confined to a three-dimensional cubic box of side length $L$ with periodic boundary conditions. The tracer and swimmer dynamics were integrated forward in time using a fourth-order Runge-Kutta method with a fixed time step $\Delta t$. The simulation algorithm proceeds as follows at each step: (i) advance swimmer, enforcing periodic wrapping. (ii) update swimmer orientation when necessary. (iii) evaluate flow field to move the tracer. 

To obtain statistically robust results, all reported quantities are ensemble averages over a large number of independent simulation runs ($N_\textmd{runs}>10000$). Each run uses a different random seed for the initial swimmer positions and for the stochastic tumble events.

In the post-processing stage, macroscopic transport coefficients were extracted from the time series of the tracer's displacement. The mean entrainment velocity was calculated from a linear fit to the mean displacement over time. The diffusion coefficients, $\mathcal{D}_\parallel$ and $\mathcal{D}_\perp$, were estimated from the long-time slope of the variance of the displacement, e.g., $2Dt=\langle \Delta z(t)^2\rangle-\langle \Delta z(t)\rangle^2$. Simulations were run for a sufficiently long duration ($T>100 \tau$) to ensure the system reached the diffusive regime where these linear relationships hold.
\bibliography{main,manual}

\end{document}